\documentclass[onecolumn,preprintnumbers,amsmath,amssymb]{revtex4}

\usepackage{graphicx}
\usepackage{epsfig}

\def\dd{\displaystyle}

\begin{document}
\title{\bf Casimir Energy Calculation for Massive Scalar Field on Spherical Surface: An Alternative Approach}

\author{M. A. Valuyan}
\email{m-valuyan@sbu.ac.ir; m.valuyan@semnaniau.ac.ir}
\affiliation{Department of Physics, Semnan Branch, Islamic Azad University, Semnan, Iran}
\date{\today}

\begin{abstract}
In this study, the Casimir energy for massive scalar field with periodic boundary condition was calculated on spherical surfaces with $S^1$, $S^2$ and $S^3$ topologies. To obtain the Casimir energy on spherical surface, the contribution of the vacuum energy of Minkowski space is usually subtracted from that of the original system. In large mass limit for surface $S^2$; however, some divergences would eventually remain in the obtained result. To remove these remaining divergences, a secondary renormalization program was manually performed. In the present work, a direct approach for calculation of the Casimir energy has been introduced. In this approach, two similar configurations were considered and then the vacuum energies of these configurations were subtracted from each other. This method provides more physical meaning respect to the other common methods. Additionally, in large mass limit for surface $S^2$, it provides a situation in which the second renormalization program is automatically conducted in the calculation procedure, and there was no need to do that anymore manually. Finally, by plotting the obtained values for the Casimir energy of the topologies and investigating their appropriate limits, the logic agreement between the results of our scheme and those of previous studies were discussed.

\end{abstract}

\maketitle

\section{Introduction}
\label{sec:intro}
\par
\begin{figure}
     \hspace{0cm}\includegraphics[width=8cm]{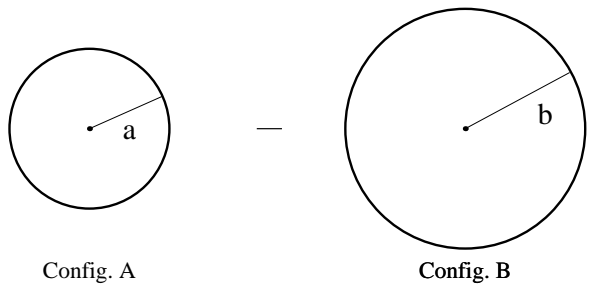}
 \caption{\label{figs.sphere}   Left (Right) figure symbolically shows a sphere with radius $a$($b$), which is named \emph{configuration A} (\emph{configuration B}). To calculate the Casimir energy, zero-point energies of these two spherical configurations should be subtracted according to  Eq.\eqref{ECasDefinition.}. At the final step, the limit $b\rightarrow\infty$, while the other parameters of problem is kept fixed. }
\end{figure}

Casimir energy is the difference between zero point energy in presence and absence of non-trivial boundary condition. This effect was first predicted and calculated by H.B.G. Casimir in 1948. He was the first one who explained the attraction between two parallel uncharged perfectly conducting plates in vacuum\,\cite{h.b.g.}. First attempts to observe this phenomenon were made in 1958 by M. J. Sparnaay\,\cite{Sparnaay.M.J.}, and then more accurate measurements verified Casimir's prediction. As Casimir effect was then considered an interesting effect of vacuum polarization, it has found many applications. This effect has an important role in different fields of physics such as quantum field theory \,\cite{quantum.field.theory.1,quantum.field.theory.2,quantum.field.theory.3,quantum.field.theory.4,quantum.field.theory.5}, condensed matter physics\,\cite{condensed.matter.1,condensed.matter.2}, atomic-molecular physics\,\cite{atomic.molecular.1,atomic.molecular.2,atomic.molecular.3,atomic.molecular.4}, gravity, astrophysics\,\cite{astro.physics.1,astro.physics.2} and mathematical physics\,\cite{Generalized.Abel.Plana.Saharian,Mathematical.physics.1,Mathematical.physics.2}. Due to the definition of the Casimir energy, two divergent terms should be subtracted from each other, which is not a simple task. In this regard, to achieve this purpose, various regularization and renormalization techniques were developed that found their own importance. In fact, Casimir energy calculations
have provided a situation where various regularization and renormalization techniques have been developed in mathematical physics. Zeta function regularization techniques\,\cite{Zeta.function.1,Zeta.function.2,Zeta.function.3}, Green's function methods\,\cite{Greens.function.} and multiple-scattering expansions\,\cite{multiple.scattering.} are some of these techniques. Advantages and disadvantages of these methods are also reviewed in previous works\,\cite{review.of.some.techniques.}. Whereas the Casimir energy is a physical quantity that its value depends on the system size (\emph{e.g.} the distance of parallel plates and radius of sphere). In some previous methods to calculate the Casimir energy, the terms, which do not depend on the system size have been eliminated from the Casimir energy expression\,\cite{wolfram.}. Physically, this elimination could be justified. However, the mathematical trend in this way cannot be maintained. In this paper, we have used a method that is free from these kinds of ambiguities in calculation of the Casimir energy. This method was first used by T.H. Boyer for the calculation of the Casimir energy for an electromagnetic field confined in a $3D$ conducting sphere\,\cite{boyer.} and it was named \emph{Box Subtraction Scheme}\,(BSS) in the later works\,\cite{BSS.1,BSS.2}. Up to now, in order to reduce possible ambiguities appearing in the calculation of the Casimir energy, multiple studies used this method\,\cite{other.BSS.1,other.BSS.2}. In the BSS, the Casimir energy is calculated by introducing two similar configurations and then their zero point energies in proper limits are subtracted from each other. To define this method concretely for our problem, as Fig.\,(\ref{figs.sphere}) shows, we consider two similar spheres. These two spheres are named A and B with radii $a$ and $b$, respectively. Then, the zero point energies of massive scalar field on the surface of these two spheres are computed and in following, the obtained vacuum energies are subtracted from each other. Finally, by taking the radius of sphere B to be infinite, the Casimir energy of sphere A will be obtained. Therefore, we have,
\begin{equation}\label{ECasDefinition.}
  \mathcal{E}_{\text{Cas.}}=\lim_{b\rightarrow\infty}\big[\mathcal{E}_A-\mathcal{E}_B\big],
\end{equation}
where $\mathcal{E}_A$ and $\mathcal{E}_B$ are zero point energy densities for $A$ and $B$ configurations, respectively. In the BSS, the contribution of vacuum energy of Minkowski space is substituted with a configuration (like sphere B) that in proper limit ($b\rightarrow \infty$), it will approach to the properties of Minkowski space. Use of two similar configurations in the BSS provides the possibility that parameters of the second configuration\,(like B configuration) play a useful role as a regulator in divergence removal. These added parameters allow the calculation of the Casimir energy via this scheme to be more clear in details. This scheme also reduces the need for using the analytic continuation techniques in the calculation process. Therefore, the related complications and ambiguities, due to analytic continuation techniques, would be avoided\,\cite{BSS.1,BSS.2,other.BSS.1,other.BSS.2}. The other successful experience of using the BSS is in higher orders of radiative corrections to the Casimir energy for different configurations, resulting in converged and consistent answers in all previous works in this category\,\cite{other.BSS.Radiative.Correction.2}. The BSS has been previously used for the calculation of the Casimir energy on flat space, but in this paper, it is intended to calculate the Casimir energy of a massive scalar field with periodic boundary condition on a curved space (\emph{e.g.} on sphere with $S^2$ and $S^3$ topologies). In common methods, to obtain the Casimir energy on curved manifolds, contribution of the vacuum energy of Minkowski space was subtracted from the vacuum energy of original manifold\,\cite{new.developements.,New.Paper.On.Sphere.,Mamaev.1979.,Advances.book.}. Since the Minkowski space is the ﬂat space, and the original manifold is curved, this subtraction is a comparing between two different kinds of spaces. The BSS, by providing the subtraction between vacuum energies of similar configurations, presents an opportunity for similar spaces to be compared with each other. In fact, the introduction of similar configurations in the BSS, in addition to creating more clarity in the computing process, also has better physical grounds. The other point in using BSS is manifested when we use it in calculation of the Casimir energy for $S^2$ topology. In all previous works, after subtracting the vacuum energy of Minkowski space from that of original system, to reach a physically consistent answer in large mass limit, an extra renormalization procedure has been performed. This secondary renormalization program to remove remaining infinities, that usually appeared due to the mass of the field, has been manually conducted\,(\emph{e.g.} Section (3.4) of Ref.\,\cite{new.developements.}). In our study, the aforementioned renormalization program is automatically performed in the BSS and there is no need to do that manually. It reflects another advantage of the BSS over curved manifolds.
\par
In the next section, through the BSS, the Casimir energy density for massive scalar field with periodic boundary condition on three spherical surface with $S^1$, $S^2$ and $S^3$ topology with radius $a$ are obtained. In the following, for each surface, the Casimir energy values in specific limits of mass of the scalar field (such as $m\rightarrow 0$ or $m \gg 1$) will be investigated. In the last section, all of our discussion concerning the BSS are summarized.

\section{The Casimir energy} \label{the Casimir energy}
In this section, the Casimir energy for massive scalar field with periodic boundary condition on surface $S^1$, $S^2$ and $S^3$ are calculated. The details of calculation for each topology is investigated in separated subsections and their extreme limits are also discussed separately. At the first step, to present a simple introduction of BSS, we start with calculation of the Casimir energy on a circle with radius $a$ .

\subsection{On a Circle ($S^1$)}

The vacuum energy density for massive scalar field with periodic boundary condition on a circle with radius $a$ can be obtained from a problem in which the massive scalar field lives in one dimension between two points with periodic boundary condition by distance $L=2\pi a$\,(for more details see Refs.\,\cite{new.developements.,other.BSS.2}). Thus, we can write the vacuum energy density of massive scalar field on a circle with radius $a$ as:
\begin{equation}\label{zero.point.1Dim.}
\mathcal{E}=\frac{1}{4\pi a}\sum_{n=-\infty}^{\infty}\omega_{n}=\frac{1}{2\pi a}\sum_{n=0}^{\infty}\omega_{n}-\frac{m}{4\pi a},
\end{equation}
where $\omega_{n}=\sqrt{m^2+\frac{n^2}{a^2}}$ is the wave number and $m$ is the mass of the field. Now, by using the definition of BSS given in Eq.\,\eqref{ECasDefinition.}, another circle with radius $b$\,($>a$), as shown in Fig.\,(\ref{figs.sphere}), is defined and the vacuum energies of these two circles should be subtracted from each other. Therefore, we have:
\begin{equation}\label{subtract.Vac.En.1Dim.}
  \mathcal{E}_{A}-\mathcal{E}_{B}=\frac{1}{4\pi a}\Big\{2\sum_{n=0}^{\infty}\sqrt{m^2+\frac{n^2}{a^2}}-m\Big\}-\{a\to b\}.
\end{equation}
All summations in Eq.\,\eqref{subtract.Vac.En.1Dim.} are divergent. Accordingly, to regularize their infinities, the Abel-Plana Summation Formula\,(APSF) is employed. This formula helps all summation forms given in Eq.\,\eqref{subtract.Vac.En.1Dim.} to be transformed to the integral form and the removal process of their infinite parts would be conducted with clarity. The usual form of APSF that we have used is:
\begin{equation}\label{Usual.APSF.}
   \sum_{n=0}^{\infty}f(n)=\frac{1}{2}f(0)+\int_{0}^{\infty}f(z)dz+ i\int_{0}^{\infty}\frac{f(it)-f(-it)}{e^{2\pi t}-1}dt.
\end{equation}
Now, by applying Eq.\,\eqref{Usual.APSF.} on the summation of Eq.\,\eqref{subtract.Vac.En.1Dim.} we have:
\begin{equation}\label{Circle.1}
  \mathcal{E}_{A}-\mathcal{E}_{B}=\Big\{\frac{m}{4\pi a}+\frac{1}{2\pi a}\int_{0}^{\infty}\sqrt{m^2+\frac{x^2}{a^2}}dx+B(a)-\frac{m}{4\pi a}\Big\}-\{a\rightarrow b\},
\end{equation}
where $\dd B(a)=\frac{-m^2}{\pi}\int_{1}^{\infty}\frac{\sqrt{z^2-1}}{e^{2\pi maz}-1}dz$ is the \emph{Branch-cut} term of APSF. The first integral term on the right hand side of Eq.\,\eqref{Circle.1} is divergent. Analogously, the same integral appears for configuration B in the second square bracket of Eq.\,\eqref{Circle.1}. To remove their infinities in subtracting procedure, we first replace the upper limit of these two integrals with multiple cutoffs $\Lambda_a$ and $\Lambda_b$ respectively. Then, by calculating integrations, we would have two separate expressions as a function of cut-offs $\Lambda_a$ and $\Lambda_b$. Now, by expanding the result in the limit $\Lambda\to\infty$, each integral becomes:
\begin{equation}\label{Integral.Expand.1Dim.}
  \int_{0}^{\Lambda}\sqrt{\frac{x^2}{a^2}+m^2}dx\buildrel {\Lambda\to\infty}\over{\longrightarrow} \frac{\Lambda^2}{2a}+\frac{m^2a}{4}-\frac{m^2a}{4}\ln(\frac{m^2a^2}{4\Lambda^2})+\mathcal{O}(\Lambda^{-2})
\end{equation}
Appropriate adjusting for cut-offs $\Lambda_a$ and $\Lambda_b$ supplemented by subtracting procedure, which is provided by BSS, helps the infinite terms appeared in the above expansion to be canceled. The BSS also helps the finite term appeared in the expansion to cancel each other out exactly. All of terms in higher order of $\mathcal{O}(\Lambda^{-2})$ do not leave any contribution in the limit $\Lambda\to\infty$. Therefore, the only remaining terms after these cancelations would be the Branch-cut terms. Finding an analytical and closed answer for integration of $B(x)$ is very cumbersome. Hence, before computation of the integral, the denominator of the integrand is expanded. Then, by calculating integral $B(x)$ we have:
\begin{equation}\label{Circle.Cas.0}
  \mathcal{E}_A-\mathcal{E}_B=\frac{-m^2}{\pi}\int_{1}^{\infty}\frac{\sqrt{z^2-1}}{e^{2\pi maz}-1}dz-\{a\to b\}=\frac{-m}{2\pi^2 a}\sum_{j=1}^{\infty}\frac{K_1(2\pi maj)}{j}-\{a\to b\}.
\end{equation}
\begin{figure}
    \hspace{0cm}\includegraphics[width=8.5cm]{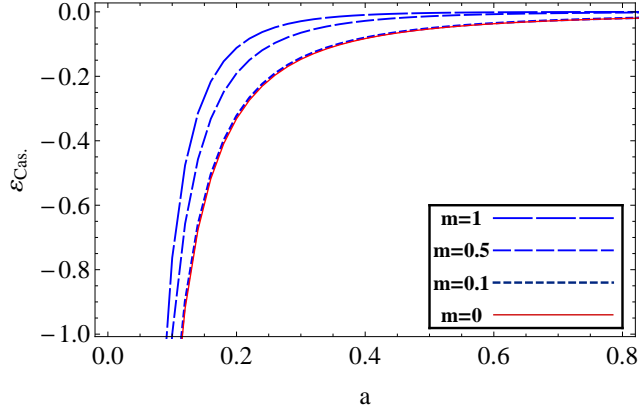}
 \caption{\label{figs.Cas.1D}   Values of the Casimir energy density for a massive scalar field on a circle are plotted as a function of radius ($a$). In this figure, we have shown a sequence of plots for $m = \{1, 0.5, 0.1, 0\}$. It is apparent that sequence of the plots for the massive cases rapidly converges into the massless case as $m$ decreases.  }
\end{figure}
Now, by applying the limit $b\to\infty$ given in Eq.\,\eqref{ECasDefinition.}, the Casimir energy density on a circle with radius $a$ becomes:
\begin{equation}\label{Circle.Cas.1}
     \mathcal{E}_{\text{Cas.}}=\frac{-\mu_a}{2\pi^2 a^2}\sum_{j=1}^{\infty}\frac{K_1(2\pi\mu_aj)}{j},
\end{equation}
where $\mu_a=ma$ and the extreme limit of the result becomes:
\begin{eqnarray}\label{extreme.limit.Circle.1}
  \hspace{-1cm}\mathcal{E}_{\text{Cas.}}\approx
  \Bigg\{\begin{array}{ll}
             \dd  \frac{-1}{24\pi a^2} \hspace{2.8cm} {\text{as}} \hspace{0.3cm} m\rightarrow0,\\
             \dd \frac{-\sqrt{\mu_a}}{4\pi^2a^2}\sum_{j=1}^{\infty}\frac{e^{-2\pi\mu_a j}}{j\sqrt{j}}  \hspace{1.1cm} {\text{as}} \hspace{0.3cm} m\gg 1.
       \end{array}
\end{eqnarray}
In Fig.\,(\ref{figs.Cas.1D}), we have plotted the Casimir energy density given in Eq.\,\eqref{Circle.Cas.1} as a function of radius $a$. This figure shows a sequence of plots for $m=\{1,0.5,0.1, 0\}$. This sequence of plots indicates that the Casimir energy for massive cases rapidly converges into the massless limit when $m\rightarrow 0$. This behavior for the Casimir energy is compatible with the previously reported results and it is also expected according to physical grounds\,\cite{new.developements.,New.Paper.On.Sphere.}.

\subsection{On a Sphere ($S^2$)}
In this subsection, we present the Casimir energy calculation via BSS for massive scalar field with periodic boundary condition on a surface with $S^2$ topology. At the first step, we remind the reader of the metric for this surface on which the scalar field lives on it. Therefore, we have:
\begin{equation}\label{metric.}
  ds^2=dt^2-a^2 (d\theta^2+\sin^2 \theta d\varphi^2 ).
\end{equation}
This metric describes $2+1$ dimensional space-time on a sphere with radius $a$ and the wave equation for a scalar field with mass $m$ on this sphere can be written as:
\begin{equation}\label{Eq.Motion.}
  \Big(\nabla_k \nabla^k+\xi \mathcal{R}+m^2\Big)\phi(\mathbf{x})=0,
\end{equation}
where $\nabla^k$ is the covariant derivative and $\mathcal{R}=2a^{-2}$ is the scalar curvature of space-time and $\xi=\frac{1}{8}$ is the conformal coupling constant. The coordinate $\mathbf{x}=(t,\theta,\varphi)$ shows the space-time coordinates on the spherical surface. By substituting the values of $\mathcal{R}$ and $\xi$ for Eq.\,\eqref{Eq.Motion.} and expanding the covariant derivative $\nabla^k$, we have:
\begin{equation}\label{Expand.Eq.Motion.}
  a^2\frac{\partial^2}{\partial t^2}\phi(\mathbf{x})-\frac{1}{\sin\theta}\frac{\partial}{\partial\theta}
  \Big(\sin\theta\frac{\partial}{\partial\theta}\phi(\mathbf{x})\Big)-
  \frac{1}{\sin^2\theta}\frac{\partial^2}{\partial \varphi^2}\phi(\mathbf{x})
  +\bigg(\frac{1}{4}+m^2a^2\bigg)\phi(\mathbf{x})=0.
\end{equation}
The orthonormal set of solutions to Eq.\,\eqref{Expand.Eq.Motion.} obeying periodic boundary conditions for both $\theta$ and $\varphi$ are represented as:
\begin{eqnarray}\label{Solution.Eq.Motion.}
 \hspace{-1cm} \Bigg\{\begin{array}{ll}
              \hspace{0cm}\phi^{(+)}_{\ell \mathcal{M}}(\mathbf{x})=\frac{1}{a\sqrt{2\omega_\ell}}e^{i\omega_\ell t}Y_{\ell \mathcal{M}}(\theta,\varphi),\\ \hspace{7cm} \ell=0,1,2,3,...\hspace{0.3cm} \mathcal{M}=0,\pm1,\pm2,...,\pm\ell \\
              \hspace{0cm}\phi^{(-)}_{\ell \mathcal{M}}(\mathbf{x})=\Big(\phi^{(+)}_{\ell \mathcal{M}}(t,\theta,\varphi)\Big)^{*}, \\
            \end{array}
\end{eqnarray}
where $Y_{\ell\mathcal{M}}(\theta,\varphi)$ are the spherical harmonic function and $\omega_\ell=\sqrt{m^2+\frac{1}{a^2} (\ell+\frac{1}{2})^2}$ are wave numbers. In order to obtain zero point energy of field on sphere, the field operator of $\phi(\mathbf{x})$ should be expanded according to the following equation:
\begin{equation}\label{Expand.Phi.}
  \phi(\mathbf{x})=\sum_{\ell=0}^{\infty}\sum_{\mathcal{M}=-\ell}^{\ell}
  \bigg[ \phi^{(-)}_{\ell \mathcal{M}}(\mathbf{x})\mathbf{a}_{\ell \mathcal{M}}+\phi^{(+)}_{\ell \mathcal{M}}(\mathbf{x})\mathbf{a}^{\dagger}_{\ell \mathcal{M}} \bigg],
\end{equation}
where $\mathbf{a}_{\ell \mathcal{M}}$ and $\mathbf{a}^{\dagger}_{\ell \mathcal{M}}$ are annihilation and creation operators of the field, respectively. The metric energy-momentum tensor is obtained by varying the Lagrangian corresponding to Eq.\,\eqref{Eq.Motion.} with respect to the metric tensor $g_{\mu\nu}$. Its diagonal $T_{00}$ component is:
\begin{eqnarray}\label{Energy.Momentun.Tensor.}
  T_{00} = (1-2\xi)\partial_0\phi\partial_0\phi+
  \Big(2\xi-\frac{1}{2}\Big)g_{00}\partial_k\phi\partial^k\phi
  -\xi\big(\phi\nabla_0\nabla_0\phi+\nabla_0\nabla_0\phi\,\phi\big) \nonumber\\
  +\bigg[\Big(\frac{1}{2}-2\xi\Big)m^2g_{00}-\xi G_{00}-2\xi^2\mathcal{R}g_{00}\bigg]\phi^2, \hspace{4.2cm}
  \end{eqnarray}
where $G_{00}=\mathcal{R}_{00}-\frac{1}{2}\mathcal{R}g_{00}$ is Einstien tensor and $\mathcal{R}_{00}$ is Ricci tensor. By substituting Eq.\eqref{Expand.Phi.} for Eq.\,\eqref{Energy.Momentun.Tensor.} and then calculating the expectation value of energy-momentum tensor, the vacuum energy density for massive scalar field on sphere will be obtained as follows:
\begin{equation}
\label{vacuum.energy}
    \mathcal{E}=<0\mid T_{00}\mid0>=\frac{1}{4\pi a^2}\sum_{\ell=0}^{\infty}\bigg(\ell+\frac{1}{2}\bigg)\omega_{\ell}.
\end{equation}
Due to the spherical symmetry, each energy level $\frac{\omega_\ell}{2}$ degenerates ($2\ell+1$) folds and high frequency modes formally render these sums as divergent.
\par
\begin{figure}
    \hspace{0cm}\includegraphics[width=8.5cm]{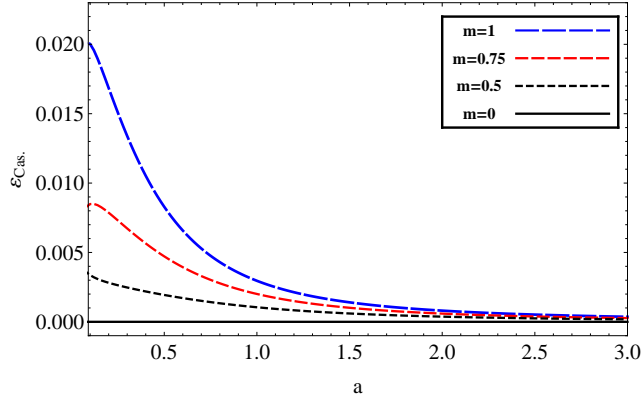}
 \caption{\label{figs.Cas.2D}   Values of the Casimir energy for a massive scalar field on a sphere with $S^2$ topology are plotted as a function of radius ($a$). In this figure, we have shown a sequence of plots for $m = \{1, 0.75, 0.5, 0\}$. It is apparent that the sequence of the plots for the massive cases rapidly converges into the massless case as $m$ decreases.  }
\end{figure}

To start the calculation of the Casimir energy via the BSS, as Fig.\,(\ref{figs.sphere}) symbolically shows, two configurations should be considered. In this problem, these configurations are two similar spheres with radii $a$ and $b$, which named \emph{A} and \emph{B} configuration, respectively. Now, by using Eqs.\,(\ref{ECasDefinition.},\ref{vacuum.energy}) we have:
\begin{equation}\label{ECas.def.substituted.}
  \mathcal{E}_{A}-\mathcal{E}_{B}=\frac{1}{4\pi a^2} \sum_{\ell=0}^{\infty}\Big(\ell+\frac{1}{2}\Big)\Big[m^2+\frac{1}{a^2}\Big(\ell+\frac{1}{2}\Big)^2\Big]^{\frac{1}{2}}
  -\{a\rightarrow b\},
\end{equation}
where $\mathcal{E}_A$ and $\mathcal{E}_B$ are zero point energy densities for $A$ and $B$ configurations, respectively. Both of the subtracted expressions in the above equation are divergent. Therefore, a regularization technique is required. At this step, the common method is using the APSF and the prescribed form of this formula for half integer parameters is\,(for a general review see Ref.\,\cite{Generalized.Abel.Plana.Saharian}):
\begin{equation}\label{APSF}
  \sum_{n=0}^{\infty}f\Big(n+\frac{1}{2}\Big)=\int_{0}^{\infty}f(z)dz-
  i\int_{0}^{\infty}\frac{f(it)-f(-it)}{e^{2\pi t}+1}dt.
\end{equation}
By applying the above form of APSF on Eq.\,\eqref{ECas.def.substituted.}, all summations are converted to integration form and become:
\begin{equation}\label{Vacuum.diff.after.APSF.}
   \mathcal{E}_{A}-\mathcal{E}_{B}=\Bigg\{\frac{1}{4\pi a^3}\int_{0}^{\infty}z\sqrt{m^2 a^2+z^2}dz+B(a)\Bigg\}-\{a\rightarrow b \},
\end{equation}
where $\dd B(x)=\frac{m^3}{2\pi}\int_{0}^{1} \frac{z\sqrt{1-z^2}}{e^{2\pi m x z}+1}dz$ is the Branch-cut term of APSF and usually has a finite value. While, the first term in both square brackets are divergent and its infinity should be removed. In order to remove infinities due to these two terms, the cutoff regularization technique is employed. Therefore, at the first step, the upper limit of integrals in Eq.\eqref{Vacuum.diff.after.APSF.} is replaced with $\Lambda_a$ and $\Lambda_b$ , respectively. Then, by calculating the integrations, we will have an answer as a function of the cutoffs $\Lambda_a$ and $\Lambda_b$, respectively. When the cutoffs $\Lambda_a$ and $\Lambda_b$ go to infinity, the following expansion for each integral is obtained:
\begin{equation}\label{expand.2Dim}
  \int_{0}^{\Lambda}z\sqrt{m^2 a^2+z^2}dz\hspace{0.1cm}\buildrel {\Lambda\to\infty}\over \longrightarrow \hspace{0.1cm} \frac{\Lambda^3}{3}+\frac{1}{2}m^2a^2\Lambda-\frac{m^3a^3}{3}
  +\mathcal{O}(\Lambda^{-1}).
\end{equation}
It can be shown that, by selecting proper values for $\Lambda_a$ and $\Lambda_b$ in the subtraction process given in Eq.\eqref{Vacuum.diff.after.APSF.}, all of the finite and infinite terms of above expansion for integral terms will be canceled and the only remaining terms in Eq.\eqref{Vacuum.diff.after.APSF.} are Branch-cut terms. Thus, we have:
\begin{equation}\label{remaining.term.}
  \mathcal{E}_{A}-\mathcal{E}_{B}=\frac{m^3}{2\pi}\int_{0}^{1}
  \frac{z\sqrt{1-z^2}}{e^{2\pi m a z}+1}dz-\{a \rightarrow b\}.
\end{equation}
Unfortunately, an analytical and closed answer for the integration of above equation does not exist. Therefore, by expanding the denominator of the integrand as the following form, we have:
\begin{equation}\label{expand.denominator.}
  \frac{1}{e^{2\pi m a z}+1}=\sum_{j=1}^{\infty}(-1)^{j+1}e^{-2\pi mazj}.
\end{equation}
After substituting Eq.\,\eqref{expand.denominator.} into Eq.\,\eqref{remaining.term.}, this becomes:
\begin{eqnarray}\label{substitude.after.expand.}
  \mathcal{E}_{A}-\mathcal{E}_{B}&=&\frac{m^3}{2\pi}\sum_{j=1}^{\infty}
  (-1)^{j+1}\int_{0}^{1}z\sqrt{1-z^2}e^{-2\pi m a z j}dz-\{a \rightarrow b\} \nonumber\\ &=&\frac{\mu_{a}^2}{8\pi a^3}\sum_{j=1}^{\infty}\frac{(-1)^{j+1}}{j}\Big[\frac{4}{3}\mu_a j-I_2(2\pi \mu_a j)+L_2(2\pi\mu_a j)\Big]-\{a \rightarrow b\},
\end{eqnarray}
where $\mu_a=ma$ and $I_2(x)$ is the modified Bessel function and $L_2 (x)$ is Struve function. At the final step, the limit $b\rightarrow\infty$ in Eq.\,\eqref{ECasDefinition.} should be computed. Thus, the Casimir energy density expression after this limit becomes:
\begin{equation}\label{Leading.order.CasEn.}
  \mathcal{E}_{\textrm{Cas.}}=\frac{\mu_{a}^2}{8\pi a^3}\sum_{j=1}^{\infty}\frac{(-1)^{j+1}}{j}\Big[\frac{4}{3}\mu_a j-I_2(2\pi \mu_a j)+L_2(2\pi\mu_a j)\Big].
\end{equation}
Our obtained expression in Eq.\,\eqref{Leading.order.CasEn.} is compatible with the previously reported result obtained in Refs.\,\cite{new.developements.,New.Paper.On.Sphere.}. The main difference between these two works is only in applying calculation methods. It seems that selecting two similar configurations and subtracting the contributions of vacuum energy of these configurations have provided a situation in which all the infinities are completely removed. It can be also shown that, the Casimir energy density for a massless scalar field vanishes. To find the large-mass limit of the Casimir energy density, we go back to the original expression for the vacuum energy density of surface, given in Eq.\,\eqref{ECas.def.substituted.}, and we select the mass $m$ as a regulator. Then, we expand the summand in the limit $m\to\infty$ and we have:
\begin{eqnarray}\label{Large.mass.limit.2D}
  \mathcal{E}_A-\mathcal{E}_B \hspace{0.1cm}\buildrel {m\to\infty}\over {\longrightarrow}\hspace{0.1cm}\frac{1}{4\pi a^2}\sum_{\ell=0}^{\infty}\Bigg[m\Big(\ell+\frac{1}{2}\Big)+\frac{1}{2ma^2}\Big(\ell+\frac{1}{2}\Big)^3+...\Bigg]-\{a\to b\}.
\end{eqnarray}
To regularize the summations, the APSF are employed. The subtraction of vacuum energies, provided by BSS, helps the infinite parts of APSF to be removed and the only remaining terms would be the Branch-cut one. Thus, we have:
\begin{eqnarray}\label{large.mass.limit.2D.2}
   \mathcal{E}_A-\mathcal{E}_B\hspace{0.1cm}\buildrel {m\to\infty}\over {\longrightarrow}\hspace{0.1cm}\frac{1}{4\pi a^2}\Big\{2m\int_{0}^{\infty}\frac{tdt}{e^{2\pi t}+1}+\frac{1}{ma^2}\int_{0}^{\infty}\frac{t^3dt}{e^{2\pi t}+1}\Big\}-\{a\to b\}.
\end{eqnarray}
In limit $m\to\infty$ the first term in the both square brackets of Eq.\,\eqref{large.mass.limit.2D.2} is still divergent. Adjusting the proper value for the parameter $m$, allows the infinities to be cancelled via BSS due to these two terms. At the last step, by computing the limit $b\to\infty$ the final remaining term for the Casimir energy in limit $m\to\infty$ is obtained as: $\mathcal{E}_{\text{Cas.}}\approx\frac{m}{48}\big(\frac{-7}{40\mu_{a}^2}\big)$.
In Fig.\,(\ref{figs.Cas.2D}), we have plotted the Casimir energy density as a function of radius $a$. In this figure, a sequence of plots for $m=\{1, 0.75, 0.5, 0\}$ is displayed. This sequence of plots shows that the Casimir energy for massive cases converges rapidly to the massless limit when $m\rightarrow 0$. This figure also shows that the Casimir energy values become zero, when the radius of sphere becomes infinite. This behavior for the Casimir energy is compatible with the previously reported results and it is also expected according to physical grounds\,\cite{new.developements.,New.Paper.On.Sphere.}.

\subsection{Three Dimensional Spherical Surface($S^3$)}
In order to find the vacuum energy density of massive scalar field on a surface with $S^3$ topology we remind the metric of the closed Friedman model as:
\begin{equation}\label{Friedmann.metric.}
   ds^2=a^2(\eta)(d\eta^2-d\chi^2-\sin^2\chi d\Omega^2),
\end{equation}
where $a(\eta)$ is a scale factor with dimension of length and $0\leq\chi\leq\pi$. Additionally, $\eta$ is a conformal time variable and $d\Omega^2=d\theta^2+\sin^2\theta d\varphi^2$. By replacing $\xi=\frac{1}{6}$ and scalar curvature $\mathcal{R}=6(a''+a)/a^3$ with Eq.\,\eqref{Eq.Motion.}, the form of equation of motion can be written as:
\begin{equation}\label{Eq.Motion.3}
  \phi''(x)+2\frac{a'}{a}\phi'(x)-\vartriangle^{(3)}\phi(x)+\Big(m^2a^2+\frac{a''}{a}+1\Big)\phi(x)=0,
\end{equation}
where $\vartriangle^{(3)}$ is the angular part of the Laplacian operator on a sphere $S^3$ and $x=(\eta,\chi,\theta,\varphi)$. To reach the vacuum energy density of system, the canonical quantization procedure should be conducted.
Hence, by solving the differential equation given in Eq.\,\eqref{Eq.Motion.3}, we have found the orthonormal set of solutions. In the following, the field operator expanded in terms of the obtained orthonormal solutions is substituted in $T_{00}$ component of stress energy-momentum tensor. Finally, the mean value of the tensor in the initial state gives us the vacuum energy density of the system as:
\begin{equation}\label{VacuumEn.3}
    \mathcal{E}=\frac{1}{4\pi^2a^4}\sum_{n=1}^{\infty}n^2\omega_n,
\end{equation}
where $\dd \omega_n=\sqrt{n^2+m^2a^2}$ is the wave number. It should be noted that for simplicity, we have followed the calculation for static Einstein model\,($a'(\eta)=0$).
\par
Now, to calculate the Casimir energy density, we have employed the BSS again. Therefore, as Fig.\,(\ref{figs.sphere}) shows, two similar spheres are considered and we define the Casimir energy by Eq.\,\eqref{ECasDefinition.}. Then, by applying the APSF given in Eq.\,\eqref{Usual.APSF.} on the summation of vacuum energy, all summations would be transformed to the integration forms and we obtain:
\begin{eqnarray}\label{ECas.3Dim}
    \mathcal{E}_A-\mathcal{E}_B=
    \frac{1}{4\pi^2a^4}\sum_{n=1}^{\infty}n^2\omega_n-\{a\rightarrow b\}=\Bigg\{\frac{1}{4\pi^2 a^4}\int_{0}^{\infty}z^2\sqrt{z^2+m^2a^2}dz+B(a)\Bigg\}-\{a\rightarrow b\}
\end{eqnarray}
where $\dd B(x)=\frac{m^4}{2\pi^2}\int_{1}^{\infty}\frac{z^2\sqrt{z^2-1}}{e^{2\pi mxz}-1}dz$ is the Branch-cut term. As is apparent, the first term in the square bracket of Eq.\,\eqref{ECas.3Dim} is divergent. To remove its infinity, the cut-off regularization scenario would be repeated the same as what occurred for Eqs.\,(\ref{Circle.1},\ref{Integral.Expand.1Dim.}). Therefore, we replace the upper limits of integrals in Eq.\,\eqref{ECas.3Dim} with a separate cutoffs $\Lambda_a$ and $\Lambda_b$, respectively. After calculating each integral and expanding their answers in limit $\Lambda\to\infty$, we have:
\begin{equation}\label{expand.int.Dim3D.}
  \int_{0}^{\Lambda}z^2\sqrt{z^2+m^2a^2}dz\buildrel {\Lambda\to\infty}\over \longrightarrow \frac{\Lambda ^4}{4}+\frac{\Lambda ^2 m^2a^2}{4}-\frac{m^4a^4}{16}\ln\left(\frac{4\Lambda^2}{m^2a^2}\right)+\frac{m^4a^4}{32}+\mathcal{O}(\Lambda^{-2}).
\end{equation}
The subtraction of vacuum energies of two configurations enable us to remove all contribution of the integral terms and only the remaining terms would be the Branch-cut one. There is not a direct way for finding of analytical and closed answer for the integral of $B(x)$. Therefore, after expanding the denominator of integrand, we calculate it. At the last step, by computing the limit $b\to\infty$, the Casimir energy density for massive scalar field on a spherical surface with $S^3$ topology is obtained as:
\begin{eqnarray}
\label{ECAS.3Dim.final}
  \mathcal{E}_{\text{Cas.}}=\frac{\mu_{a}^2}{8\pi^4a^4}\sum_{j=1}^{\infty}\frac{2\pi\mu_a j K_1(2\pi \mu_a j)+3K_2(2\pi\mu_a j)}{j^2},
\end{eqnarray}
where $\mu_a=ma$ and the main following limits for this result are:
\begin{eqnarray} \label{ECAS.3Dim.limit.}
\hspace{-1cm}\mathcal{E}_{\text{Cas.}}\approx
\Bigg\{\begin{array}{ll}
               \dd\frac{1}{480\pi^2a^4} \hspace{2.3cm} {\text{as}} \hspace{0.5cm} m\rightarrow0,\\
               \dd\frac{\mu_{a}^2\sqrt{\mu_a}}{8\pi^3a^4}\sum_{j=1}^{\infty}\frac{e^{-2\pi\mu_a j}}{j^{\frac{3}{2}}}  \hspace{1.1cm} {\text{as}} \hspace{0.3cm} m\gg 1.
       \end{array}
\end{eqnarray}
\begin{figure}
    \hspace{0cm}\includegraphics[width=8.5cm]{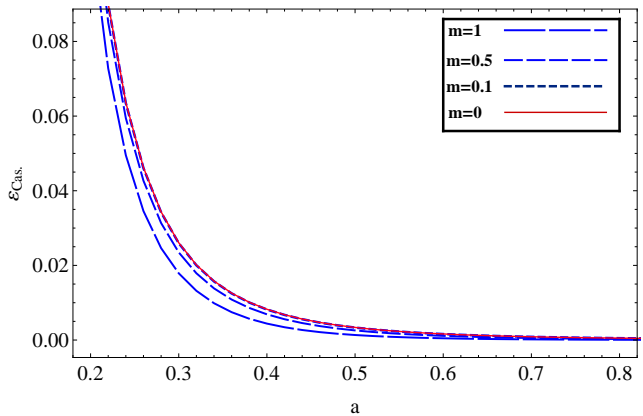}
 \caption{\label{figs.Cas.3D}   Values of the Casimir energy density for a massive scalar field on a sphere with $S^3$ topology are plotted as a function of radius ($a$). In this figure, we have shown a sequence of plots for $m = \{1, 0.5, 0.1, 0\}$. It is apparent that the sequence of the plots for the massive cases rapidly converges into the massless case as $m$ decreases.  }
\end{figure}
In Fig.\,(\ref{figs.Cas.3D}), the Casimir energy density as a function of radius $a$ for a sequence of plots of $m=\{1,0.5,0.1, 0\}$ is displayed. This plot shows that the Casimir energy for massive cases rapidly converges into the massless limit when $m\rightarrow 0$. This behavior for the Casimir energy is compatible with the previously reported results\,\cite{new.developements.,New.Paper.On.Sphere.}. The very simple elimination of divergences in BSS shows that this method is highly powerful in this task and it can be a good candidate in removal divergences in the known regularization techniques, specially for curved spaces. This method, by comparing two similar kinds of manifolds in its definition, provides sufficient degrees of freedom to adjust the cutoffs for removal process and it conceptually provides more physical grounds.

\section{Conclusions}
\label{sec:conclusion}
In the present paper, the Casimir energy for the massive scalar field with periodic boundary condition on the spherical surface with $S^1$, $S^2$ and $S^3$ topologies were calculated. In the present calculation, procedure of the Box Subtraction Scheme (BSS) as an alternative and direct method was introduced and its various aspects were discussed. In this scheme, the Casimir energy was obtained by subtracting the vacuum energy of two similar configurations in proper limits. This subtraction in the BSS enables us to remove all divergences clearly. We maintain that subtracting the vacuum energy of a curved manifold with nontrivial topology from the vacuum energy of flat space\,(\emph{e.g.} the Minkowski space) is irrelevant and at least has less physical meaning. Indeed, in that subtraction, two different kinds of spaces are compared with each other. In the BSS, we allow two similar kinds of vacuum energies to be compared with each other and we maintain that, this subtraction has a more physical ground than the other common definitions of the Casimir energy. The BSS also provides a directive way in calculation of the Casimir energy on spherical surface with $S^2$ topology in proper limit. The obtained results are consistent with expected physical grounds and also in good agreement with the previously reported results. As applying some regularization techniques like analytical continuation technique, give rise to some ambiguities in the Casimir energy calculation\,\cite{other.BSS.Radiative.Correction.2}, it is hoped that the BSS as a regularization technique will be successful in reducing these sorts of ambiguities. It is also anticipated that the mentioned scheme will be useful in radiative correction to the Casimir energy in various boundary conditions on curved manifolds.

\acknowledgments
The Author would like to thank the research office of Semnan Branch, Islamic Azad University for the financial support.


\begin{thebibliography}{99}

\bibitem{h.b.g.}
H. B. G. Casimir, Proc. Kon. Nederl. Akad. Wet. {\bf 51} (1948) 793.

\bibitem{Sparnaay.M.J.}
     M. J. Sparnaay,\emph{Measurements of attractive forces between flat plates}, Physica {\bf 24} (1958) 751.

\bibitem{quantum.field.theory.1}
     K. A. Milton, \emph{The Casimir Effect: Physical Manifestations of Zero-Point Energy}, (World
     Scientific Publishing Co. 2001).
\bibitem{quantum.field.theory.2}
     M. Bordag, E. Elizalde, K. Kirsten and S. Leseduarte, Phys. Rev. D {\bf 56} (1997) 4896.

\bibitem{quantum.field.theory.3}
     V. N. Marachevsky, \emph{Casimir Effect For Periodic Systems Separated by a Vacuum Gap}, Theo. Math. Phys. \textbf{176} (2013) 929.

\bibitem{quantum.field.theory.4}
     N. R. Khusnutdinov and R. N. Kashapov, \emph{Casimir Effect For a Colection of Parallel Conducting Surfaces}, Theo. Math. Phys.  {\bf183} (2015) 491.
\bibitem{quantum.field.theory.5}
     C. K. Chewand and  R. T. Sharp, \emph{On The Degeneracy Problem in SU(3)}, Can. J. Phys. \text{44} (1966) 2789.

\bibitem{condensed.matter.1}
    D. C. Roberts and Y. Pomeau, Phys. Rev. Lett. \textbf{95} (2005) 145303.
\bibitem{condensed.matter.2}
    A. Edery, Phys. Rev. D \textbf{75} (2007) 105012.


\bibitem{atomic.molecular.1}
    M. Krech and S. Dietrich, Phys. Rev. Lett. \textbf{66} (1991) 345.
\bibitem{atomic.molecular.2}
    M. A. Braun, \emph{Casimir Energy of The Quantum Field in a Dispersive and Absorptive Medium}, Theo. Math. Phys. {\bf175} (2013) 771.
\bibitem{atomic.molecular.3}
    I. Brevik and H. Kolbenstvedt, \emph{Casimir stress in spherical media when $\epsilon\mu= 1$}, Can. J. Phys. \textbf{62} (1984) 805.
\bibitem{atomic.molecular.4}
    I. Brevik and H. Kolbenstvedt, \emph{Attractive Casimir stress on a thin spherical shell}, Can. J. Phys. \textbf{63} (1985) 1409.


\bibitem{astro.physics.1}
    G. Mahajan, S. Sarkar and T. Padmanabhan, Phys. Lett. B \textbf{641} (2006) 6.

\bibitem{astro.physics.2}
    H. Kleinert, A. Zhuk, \emph{The Casimir effect at nonzero temperatures in a Universe with topology $S^1 \times S^1 \times S^1$}, Theo. Math. Phys. {\bf 108} (1996) 482.


\bibitem{Generalized.Abel.Plana.Saharian}
     A.A. Saharian, \emph{The Genetalized Abel-Plana Formula: Applications To Bessel Functions And Casimir Effect} IC/2007/082 (2000)
     [{ \tt hep-th/0002239 v1}].

\bibitem{Mathematical.physics.1}
     A. Romeo, K.A. Milton, \emph{Casimir energy for a purely dielectric cylinder by the mode summation method}, Phys. Lett. B \textbf{621} (2005) 309.

\bibitem{Mathematical.physics.2}
     E. M. Santangelo, \emph{Evaluation of Casimir Energies Through Spectral Functions}, Theo. Math. Phys. {\bf 131} (2002) 98.

\bibitem{Zeta.function.1}
     K.A. Milton, A.V. Nesterenko, V.V. Nesterenko, \emph{Mode-by-mode summation for the zero point electromagnetic energy of an infinite cylinder}, Phys. Rev. D \textbf{59} (1999) 105009.
\bibitem{Zeta.function.2}
     I.H. Brevik, V.V. Nesterenko, I.G. Pirozhenko, \emph{Direct mode summation for the Casimir energy of a solid ball}, J. Phys. A \textbf{31} (1998) 8661.
\bibitem{Zeta.function.3}
     V.V. Nesterenko, I.G. Pirozhenko, \emph{Spectral zeta functions for a cylinder and a circle}, J. Math. Phys. \textbf{41} (2000) 4521.

\bibitem{Greens.function.}
     K.A. Milton, L.L. Deraad, J. Schwinger, Casimir self-stress on a perfectly conducting spherical shell, Ann. Phys. (N.Y.) \textbf{115} (1978) 388.

\bibitem{multiple.scattering.}
     R. Balian, B. Duplantier, \emph{Electromagnetic waves near perfect conductors. II. Casimir effect}, Ann. Phys. (N.Y.) \textbf{112} (1978) 165.


\bibitem{review.of.some.techniques.}
     M. Bordag and K. Kiresten, \emph{Heat kernel Coefficients and Divergencies of the Casimir Energy for the Dispersive Sphere}, Int. J. Mod. Phys. A \textbf{17} (2002) 813.
\bibitem{wolfram.}
     J. Ambj{\o}rn and S. Wolfram, 1983 \emph{Properties of the vacuum: I. Mechanical and thermodynamic} Ann. Phys., NY \textbf{147}  (1983) 1.

\bibitem{boyer.}
     T. H. Boyer, Phys. Rev. \textbf{174} (1968) 1764 (1968).

\bibitem{BSS.1}
     M.A. Valuyan, R. Moazzemi, S.S. Gousheh, \emph{A direct approach to the electromagnetic Casimir energy in a rectangular waveguide}, J. Phys. B: At. Mol. Opt. Phys. \textbf{41} (2008) 145502.
\bibitem{BSS.2}
     M.A. Valuyan, S.S. Gousheh, \emph{Dirichlet Casimir Energy For a Scalar Field in a Sphere: An Alternative Method}, Int. J. Mod. Phys. A \textbf{25} (2010) 1165.
\bibitem{other.BSS.1}
    R. Moazzemi, A. Mohammadi, S. S. Gousheh, Eur. Phys. J. C {\bf 56} (2008) 585.
\bibitem{other.BSS.2}
    R. Moazzemi, S. S. Gousheh, Phys. Lett. B {\bf 658} (2008) 255.
\bibitem{other.BSS.Radiative.Correction.2}
    S. S. Gousheh, R. Moazzemi and  M. A. Valuyan, Phys. Lett. B \textbf{681} (2009) 477.

\bibitem{new.developements.}
     M. Bordag, U. Mohideen, and V. M. Mostepanenko, \emph{New developments in the Casimir effect}, Phys. Rep. \textbf{353} (2001) 1. [{\tt arXiv:quant-ph/0106045}].

\bibitem{New.Paper.On.Sphere.}
     C. A. R. Herdeiro, R. H. Ribeiro and M. Sampaio, Class. Quantum Grav. \textbf{25} (2008)165010.

\bibitem{Mamaev.1979.}
    S. G. Mamaev and N. N.Trunov, \emph{Vacuum Means of Energy-Momentum Tensor of Quantized Fields on Manifolds of Different Topology and Geometry}, Soviet Phys. J. {\bf 22} (1979) 766.

\bibitem{Advances.book.}
    M. Bordag, G. L.Klimchitskaya, U. Mohideen, V. M. Mostepanenko, \emph{Advances in the Casimir Effect}, Oxford Publishing, (2009).

\end{thebibliography}
\end{document}